\documentclass[preprint,aps,epsfig]{revtex4}

\usepackage{psfig}

\def\lsim{\raise0.3ex\hbox{$<$\kern-0.75em\raise-1.1ex\hbox{$\sim$}}}
\def\gsim{\raise0.3ex\hbox{$>$\kern-0.75em\raise-1.1ex\hbox{$\sim$}}}

\def\pom{{I\!\!P}}
\newcommand{\be}{\begin{equation}}
\newcommand{\ee}{\end{equation}}

\def\beq{\begin{equation}}
\def\eeq{\end{equation}}
\def\beqa{\begin{eqnarray}}
\def\eeqa{\end{eqnarray}}

\newcommand{\rr}{\mbox{\boldmath $r$}}
\newcommand{\rrn}{\mbox{$r$}}

\newcommand{\rb}{\mbox{\boldmath $b$}}

\def\gappeq{\mathrel{\rlap {\raise.5ex\hbox{$>$}}
{\lower.5ex\hbox{$\sim$}}}}

\def\lappeq{\mathrel{\rlap{\raise.5ex\hbox{$<$}}
{\lower.5ex\hbox{$\sim$}}}}

\def\Toprel#1\over#2{\mathrel{\mathop{#2}\limits^{#1}}}

\def\pom{{I\!\!P}}

\begin{document}

\title{Saturation in diffractive deep inelastic $eA$ scattering} 
\author{  M.S. Kugeratski$^1$, V.P. Gon\c{c}alves$^2$, and  F.S. Navarra$^1$}
\affiliation{$^1$Instituto de F\'{\i}sica, Universidade de S\~{a}o Paulo, 
C.P. 66318,  05315-970 S\~{a}o Paulo, SP, Brazil\\
$^2$High and Medium Energy Group (GAME), \\
Instituto de F\'{\i}sica e Matem\'atica,  Universidade
Federal de Pelotas\\
Caixa Postal 354, CEP 96010-900, Pelotas, RS, Brazil\\}
\begin{abstract}
In this paper we  investigate the saturation physics in diffractive deep inelastic electron-ion scattering. 
We estimate the energy and nuclear dependence of the ratio $\sigma^{diff}/\sigma^{tot}$  and predict the 
$x_{\pom}$ and $\beta$ behavior of the nuclear diffractive structure function 
$F_{2,A}^{D(3)}(Q^2, \beta, x_{I\!\!P})$. Moreover, we analyze the ratio 
 $R^{diff}_{A1,A2}(Q^2, \beta, x_{I\!\!P}) = F_{2,A1} ^{D(3)}/F_{2,A2} ^{D(3)}$, which probes the nuclear dependence of the structure of the Pomeron.
We show that saturation physics predicts that approximately 37 $\%$ of the events observed at eRHIC should be diffractive.

\end{abstract}
\maketitle
\vspace{1cm}

\section{Introduction}

One of the  frontiers of QCD which are intensely investigated in high energy experiments 
is the high energy (small $x$) regime, where we expect to observe the non-linear behavior of the theory. 
In this regime, the growth of the parton distribution should saturate, forming a  Color Glass Condensate (CGC). 
 (For recent reviews see, e.g. Refs. \cite{iancu_raju,anna,weigert,JK}). In fact, signals of parton saturation have already
been observed both in  $ep$ deep inelastic scattering at HERA and in deuteron-gold 
collisions at RHIC (See, e.g. Ref. \cite{blaizot}). However, the observation of this new regime still 
needs confirmation and so there is an active search for new experimental signatures. 
Among them, the observables measured in diffractive deep inelastic scattering (DDIS)  deserve special 
attention. As shown in Ref. \cite{GBW}, the total diffractive cross section is much more sensitive to 
large-size dipoles than the inclusive one. Saturation effects  screen large-size dipole (soft) contributions, so that a fairly large 
fraction of the cross section is hard and hence eligible for a perturbative treatment. Therefore,  the study of diffractive processes 
becomes fundamental in order to constrain the QCD dynamics at high energies.

Significant progress in understanding diffraction has been made at the $ep$ collider HERA 
(See, e.g. Refs. \cite{MW,H,PREDAZZI}). Currently, there exist many attempts to describe the diffractive part of 
the deep inelastic cross section within  pQCD (See, e.g. Refs. \cite{GBW,fss,MRW,Brod}). One of the most successful 
approaches is the saturation one \cite{GBW}  based  on the dipole picture of DIS \cite{dipole,dipole2}. It naturally 
incorporates the description of both inclusive and diffractive events in a common theoretical framework, as the 
same dipole scattering amplitude enters in the formulation of the inclusive and diffractive cross sections. 
In the studies of saturation effects in DDIS,  non-linear evolution equations for the dipole scattering
 amplitude have been derived \cite{BK,KL,KW,W}, new measurements proposed \cite{slopedif1,slopedif2,MS,GM} 
 and the charm contribution estimated \cite{charmdif}. However, as shown in Ref. \cite{fss}, current data are not yet 
 precise enough, nor do they extend to sufficiently small values of $x_{\pom}$, to discriminate between different 
 theoretical approaches.

Other source of information on QCD dynamics at high parton density is due to nuclei which provide 
high density at comparatively lower energies. Recently, in Ref. \cite{eA}, we have estimated a set of 
inclusive observables which could be analyzed in a future electron-ion collider \cite{raju_eRHIC}. Our results have demonstrated 
that the saturation physics cannot be disregarded in the kinematical range of eRHIC. Our goal in this work is to 
understand to what extend the saturation regime of QCD manifests itself in diffractive deep inelastic $eA$ scattering. 
In particular, we will study the energy and nuclear dependence  of the ratio between  diffractive and 
total cross sections ($\sigma_{diff}/\sigma_{tot}$). HERA has observed that the energy dependence of this 
ratio is almost constant for different mass intervals of the diffractively produced hadrons over a wide range of 
photon virtualities $Q^2$ \cite{e_p_data}. This ratio is to a good approximation constant as a function of the Bjorken 
$x$ variable and $Q^2$. Moreover, we make predictions for more detailed diffractive properties, such as 
those embodied in the diffractive structure function $F_2^{D (3)} (Q^2, \beta, x_{\pom})$. Motivated by Refs. 
\cite{Arneodoetal,raju_eRHIC} we also analyze the behavior of the ratio between nuclear diffractive structure 
functions $R^{diff}_{A1,A2}( Q^2, \beta, x_{I\!\!P}) = F_{2,A1} ^{D(3)}/F_{2,A2} ^{D(3)}$, where $A_1$ and $A_2$ denote 
the atomic number of the two nuclei. It is important to emphasize that diffractive processes in $eA$ collisions were 
studied in Refs. \cite{nzz,KM,gllmt,glmmt,levlub,fgs04,MS, Goncalves2,Goncalves3}. Here we extend these studies  for a large number of 
observables, considering the dipole approach and  a generalization for nuclear targets  of the CGC dipole cross 
section proposed in Ref. \cite{IIM}. As this model  successfully describes the HERA data, we believe that it is 
possible to obtain realistic predictions for the kinematical range of the electron-ion collider eRHIC.


 This paper is organized as follows. In next Section we present a brief review of the dipole picture. We present 
 the main formulae for the dipole cross section and the diffractive structure function. 
In Section \ref{res} we introduce  the overlap function for diffractive events, which allows us to find out the average
 dipole size which contributes the most to this process, and analyze its dipole size and nuclear dependences. Moreover, 
 we estimate the different contributions to the diffractive structure function and present our predictions for $F_2^{D (3)}$ and  $R^{diff}_{A1,A2}$. Finally, in Section \ref{conc} we summarize our main conclusions.

\section{Dipole picture of diffractive DIS}
 
In deep inelastic scattering, a photon of virtuality $Q^2$ collides
with a target. In an appropriate frame, called the dipole frame, the virtual 
photon undergoes the hadronic interaction via a fluctuation into a dipole. The  
wavefunctions $|\psi _T|^2$ and $|\psi _L|^2$, describing the 
splitting of the photon 
on the dipole, are given by ~\cite{dipole} :
\begin{eqnarray}
\label{eq:psi^2}
|\psi_{L}(\alpha,r)|^{2} & =  & \frac{3 \alpha_{em}}{\pi^{2}}\sum_f e_{f}^
{2} 4 Q^{2}\alpha^{2}(1-\alpha)^{2} K_{0}^{2}(\epsilon r) \\
  |\psi_{T}(\alpha,r)|^{2} & = & \frac{3 \alpha_{em}}{2 \pi^{2}}\sum_{f} e_{f}^
{2} \left\{[\alpha^{2} + (1-\alpha)^{2}] \epsilon^{2} K_{1}^{2}(\epsilon r) + m_{f}^{2} 
K_{0}^{2}(\epsilon r) \right\} 
\end{eqnarray}
for a longitudinally and transversely   polarized photon,  respectively. In the above 
expressions  
\(
 \epsilon^{2} = \alpha(1-\alpha)Q^{2} + m_{f}^{2}\; ,
\) 
$K_{0}$ and $K_{1}$ are modified Bessel functions and the sum is over quarks of flavor 
$f$  with  a corresponding  quark mass $m_f$. As usual $\alpha$ stands for  the 
longitudinal photon momentum fraction carried by the quark and $1 - \alpha$ is the 
longitudinal photon momentum fraction of  the antiquark. The dipole then 
interacts with the target  and one has the following factorized formula 
\cite{dipole}
\begin{equation}
\sigma^{\gamma^{*} A} _{L,T} (x,Q^2) = \int d\alpha \;  d^2 \rr \ 
|\psi_{L,T}(\alpha,\rr)|^{2} \sigma_{dip}(x,\rr) \,\,.
\label{sig_gp}  
\end{equation} 
Similarly, 
the total diffractive cross sections take on the following form  (See e.g. Refs. \cite{GBW,PREDAZZI,dipole})
\begin{equation}
\sigma^D_{T,L} = \int_{-\infty}^0 dt\,e^{B_D t} \left. \frac{d \sigma ^D _{T,L}}{d t} \right|_{t = 0} = \frac{1}{B_D} \left. \frac{d \sigma ^D _{T,L}}{d t} \right|_{t = 0}
\label{sig_difra}
\end{equation}
where 
\begin{equation}
\left. \frac{d \sigma ^D _{T,L}}{d t} \right|_{t = 0} = \frac{1}{16 \pi} \int d^2 {\bf r} 
\int ^1 _0 d \alpha |\Psi _{T,L} (\alpha, {\bf r})|^2 \sigma _{dip} ^2 (x, r^2) \,\,,
\end{equation}
and we have assumed a factorizable dependence on $t$ with diffractive slope $B_D$.

The diffractive process can be analyzed in more detail studying the behavior of the diffractive structure function $F_2^{D (3)}(Q^{2}, \beta, x_{I\!\!P})$. In Refs. \cite{GBW,dipole} the authors have derived expressions for $F_2^{D (3)}$ directly in the transverse momentum space and then transformed to impact parameter space where the dipole approach can be applied. Following Ref. \cite{GBW} we assume that the diffractive structure function is given by
\begin{equation}
F_2^{D (3)} (Q^{2}, \beta, x_{I\!\!P}) = F^{D}_{q\bar{q},L} + F^{D}_{q\bar{q},T} + F^{D}_{q\bar{q}g,T}
\end{equation}
where $T$ and $L$ refer to the polarization of the virtual photon. For the $q\bar{q}g$ contribution only the transverse polarization is considered, since the longitudinal counterpart has no leading logarithm in $Q^2$.  The computation of the different contributions was made in Refs. \cite{wusthoff,GBW,dipole} and here we quote only the final results:
\begin{equation}
  x_{I\!\!P}F^{D}_{q\bar{q},L}(Q^{2}, \beta, x_{I\!\!P})=
\frac{3 Q^{6}}{32 \pi^{4} \beta B_D} \sum_{f} e_{f}^{2} 
 2\int_{\alpha_{0}}^{1/2} d\alpha \alpha^{3}(1-\alpha)^{3} \Phi_{0},
\label{qqbl}
\end{equation}
\begin{equation}
 x_{I\!\!P}F^{D}_{q\bar{q},T}(Q^{2}, \beta, x_{I\!\!P}) =  
 \frac{3 Q^{4}}{128\pi^{4} \beta B_D}  \sum_{f} e_{f}^{2} 
 2\int_{\alpha_{0}}^{1/2} d\alpha \alpha(1-\alpha) 
\left\{ \epsilon^{2}[\alpha^{2} + (1-\alpha)^{2}] \Phi_{1} + m_f^{2} \Phi_{0}  \right\}   
\label{qqbt}
\end{equation}
where the lower limit of the integral over $\alpha$ is given by $\alpha_{0} = \frac{1}{2} \, \left(1 - \sqrt{1 - \frac{4m_{f}^{2}}{M_X^{2}}}\right)
$ and we have introduced the auxiliary functions \cite{fss}:
\begin{equation}
\Phi_{0,1}  \equiv  \left(\int_{0}^{\infty}r dr K_{0 ,1}(\epsilon r)\sigma_{dip}(x_{I\!\!P},r) J_{0 ,1}(kr) \right)^2.
\label{fi}
\end{equation}
 For the $q\bar{q}g$ contribution  we have \cite{wusthoff,GBW,nikqqg}
 \begin{eqnarray}
   \lefteqn{x_{I\!\!P}F^{D}_{q\bar{q}g,T}(Q^{2}, \beta, x_{I\!\!P}) 
  =  \frac{81 \beta \alpha_{S} }{512 \pi^{5} B_D} \sum_{f} e_{f}^{2} 
 \int_{\beta}^{1}\frac{\mbox{d}z}{(1 - z)^{3}} 
 \left[ \left(1- \frac{\beta}{z}\right)^{2} +  \left(\frac{\beta}{z}\right)^{2} \right] } \label{qqg} \\
  & \times & \int_{0}^{(1-z)Q^{2}}\mbox{d} k_{t}^{2} \ln \left(\frac{(1-z)Q^{2}}{k_{t}^{2}}\right) 
\left[ \int_{0}^{\infty} u \mbox{d}u \; \sigma_{dip}(u / k_{t}, x_{I\!\!P}) 
   K_{2}\left( \sqrt{\frac{z}{1-z} u^{2}}\right)  J_{2}(u) \right]^{2}.\nonumber
\end{eqnarray} 
We  use the standard notation for the variables $\beta = Q^2 / (M_X^2 + Q^2)$, $
x_{I\!\!P} = (M_X^2 + Q^2)/(W^2 + Q^2)$ and $x = Q^2/(W^2 + Q^2) = \beta x_{\pom}$, 
where $M_X$ is the invariant mass of the diffractive system and $W$ the total energy of the 
$\gamma ^* p$ (or $\gamma ^* A$ ). When  extending (\ref{qqbl}),  (\ref{qqbt}) and (\ref{qqg}) to the nuclear case  we need to change 
the slope to the nuclear slope parameter, $B_A$. In what follows we will assume that $B_A$ may be approximated by $B_A = \frac{R_A ^2}{4}$, where $R_A$ is given by
$R_A = 1.2 A^{1/3}$ fm  ~\cite{frankfurt}.

In the dipole picture  the behavior of the total inclusive and diffractive cross sections, as well as the diffractive 
structure functions, is strongly dependent on the dipole cross section, which is determined by the QCD dynamics. Consequently, in the dipole picture the inclusion of saturation physics is quite transparent and 
straightforward, as the dipole cross section is closely related to the solution of the 
QCD non-linear evolution equations (For recent reviews see, e.g. Refs. \cite{iancu_raju,anna,weigert,JK})
\begin{eqnarray}
\sigma_{dip} (x,\rr)=2 \int d^2 \rb \, {\cal{N}}(x,\rr,\rb)\,\,,
\end{eqnarray}
where ${\cal{N}}$ is the  quark dipole-target forward scattering amplitude for a given impact 
parameter $\rb$  which encodes all the
information about the hadronic scattering, and thus about the
non-linear and quantum effects in the hadron wave function. 
In what follows we will disregard the impact parameter dependence [$\sigma_{dip} = \sigma_0 \, {\cal{N}}(x,\rr)$] 
and  consider the phenomenological model proposed in Ref. \cite{IIM}, in which    a parameterization of  
${\cal{N}} (x,\rr)$ was constructed so as to reproduce two limits  
analytically under control: the solution of the BFKL equation
for small dipole sizes, $\rr\ll 1/Q_s(x)$, and the Levin-Tuchin law \cite{Levin}
for larger ones, $\rr\gg 1/Q_s(x)$. 
Here, $Q_s$ denotes the saturation momentum scale, which is the basic quantity characterizing the saturation effects, 
being related to a critical transverse size for the unitarization of the cross section, and is an increasing function of  the energy [$Q_s^2 =  Q_0^2 \, (\frac{x_0}{x})^{\lambda}$].
A fit to the structure function $F_2(x,Q^2)$ 
was performed in the kinematical range of interest, showing that it is  not very 
sensitive to the details of the interpolation. The  dipole-target forward scattering amplitude  was parameterized 
as follows,
\begin{eqnarray}
{\cal{N}}(x,\rr) =  \left\{ \begin{array}{ll} 
{\mathcal N}_0\, \left(\frac{\rr\, Q_s}{2}\right)^{2\left(\gamma_s + 
\frac{\ln (2/\rr Q_s)}{\kappa \,\lambda \,Y}\right)}\,, & \mbox{for $\rr Q_s(x) \le 2$}\,,\\
 1 - \exp^{-a\,\ln^2\,(b\,\rr\, Q_s)}\,,  & \mbox{for $\rr Q_s(x)  > 2$}\,, 
\end{array} \right.
\label{CGCfit}
\end{eqnarray}
where $a$ and $b$ are determined by continuity conditions at $\rr Q_s(x)=2$, $\gamma_s= 0.63$, $\kappa= 9.9$, $\lambda=0.253$, $Q_0^2 = 1.0$ GeV$^2$,
$x_0=0.267\times 10^{-4}$ and ${\mathcal N}_0=0.7$. Hereafter, 
we label the model above by IIM.
The first line from Eq. (\ref{CGCfit}) describes the linear regime whereas the 
second one describes saturation effects. When estimating the relative importance 
of saturation we will switch off the second line of (\ref{CGCfit}) and use only 
the first. This is a relevant check, since some observables may turn out to be completely 
insensitive to non-linear effects. 
Following Ref. \cite{eA},  we generalize the IIM  model for nuclear collisions  assuming  the following basic transformations: $\sigma_0 \rightarrow \sigma_0^A =  A^{\frac{2}{3}} \times \sigma_0$ and $Q_s^2(x) \rightarrow Q_{s,A}^2 =  A^{\frac{1}{3}} \times  Q_s^2(x)$. As already emphasized in that reference, more sophisticated generalizations
 for the nuclear case are possible. However, as our goal is to obtain a first estimate of the saturation effects in these 
 processes, our choice was to consider a simplified model which introduces a minimal set of assumptions. In a full 
calculation we must  use the solution of the BK equation, obtained without disregarding the impact parameter 
dependence as well as an initial condition constrained by current experimental data on lepton-nucleus DIS.

\begin{figure}
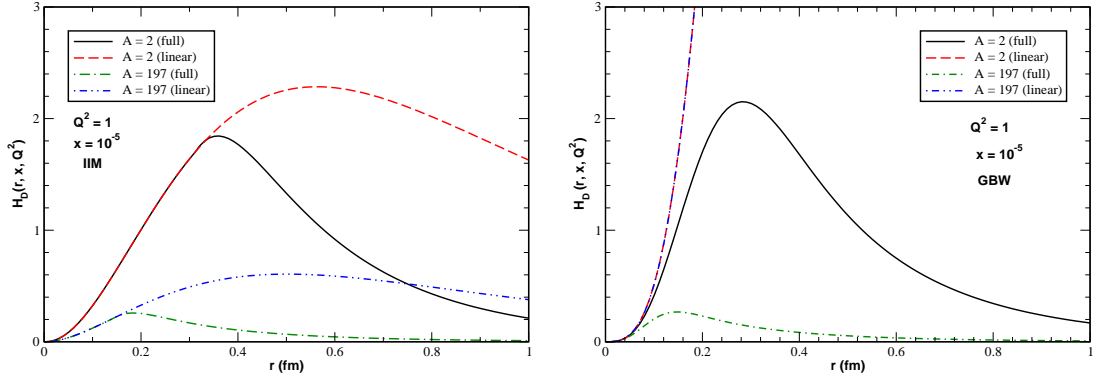

\centerline{
\begin{tabular}{ccc}
{\psfig{figure=fig1a.eps,width=7.cm}} & \,\,\, {\psfig{figure=fig1b.eps,width=7.cm}} &  \\
 &  & \\
\end{tabular}}
\caption{The $\rr$-dependence of the photon-nucleus diffractive overlap function (normalized by $A^2$) at different values of the atomic number and distinct saturation models: IIM (left panel) and GBW (right panel).}
\label{fig1}
\end{figure}

\section{Results}
\label{res}

Before presenting our results for $\frac{\sigma_{diff}}{\sigma_{tot}}$ and for the diffractive structure function, let's  
investigate the mean dipole size dominating the diffractive cross section through the analysis of the  photon-nucleus 
diffractive overlap function defined by 
\begin{eqnarray}
H_D \,(\rr,x,Q^2)  =  2\pi \rrn \,\sum_{i=T,L} \int d \alpha \,  |\Psi_i
(\alpha,\,\rr, Q^2)|^2 \, \sigma_{dip}^2 (x,\rr,A)\,.
\label{overlap}
\end{eqnarray}
In Fig. \ref{fig1} we present the $\rr$-dependence of the photon-nucleus diffractive overlap function (normalized by $A^2$) at different 
values of the atomic number and $Q^2 = 1$ GeV$^2$. A similar analysis can be made for other values of $Q^2$. The main 
difference is that at large values of $Q^2$, the overlap function peaks at smaller values of the pair separation. 
For comparison the predictions obtained with a 
generalization of the Golec-Biernat W\"usthoff  (GBW) model \cite{GBW} for nuclear targets is also presented. When the 
saturation effects are included the IIM and GBW overlap functions present 
a similar behavior, strongly reducing the contribution of large pair separations. At large $A$ only small pair separations contribute to
 the diffractive cross section.
 However, in the linear case, these two models present a very distinct behavior, 
which is directly associated with the different prescription for the linear regime. While the GBW model assumes 
that $\sigma_{dip} \propto \sigma_0 \rr^2 Q_s^2$ in this regime, the IIM one assumes 
$\sigma_{dip} \propto \sigma_0 [\rr^2 Q_s^2]^{\gamma_{eff}}$, where $\gamma_{eff} = \gamma_s + 
\ln (2/\rr Q_s)/\kappa \,\lambda \,Y$ is smaller than one. Firstly, it implies a different $A$ dependence, since 
in the GBW model the product $[\sigma_0 Q_s^2]^2$ is proportional to $A^2$ which cancels with the normalization term.
In the IIM model we have $[\sigma_0 Q_s^{2 \gamma_{eff}}]^2$, which implies an $A^{\frac{4 + 2 \gamma_{eff}}{3}}$ dependence. 
When combined with the factor $A^2$ which comes from the normalization, we expect an 
$A^{\frac{ 2 (\gamma_{eff} - 1) }{3}}$ dependence for the IIM overlap function. As $\gamma_{eff} < 1$, the 
overlap function decreases for large $A$ also in the linear regime. This behavior is seen in Fig. \ref{fig1}. 
Secondly, in contrast to the GBW model which predicts a $\rr^2$ behavior for the dipole cross section in the 
linear regime, the IIM one leads to a $\rr^{2\gamma_{eff}}$ dependence. This different prescription for the $\rr$-dependence, when combined with the pair separation dependence of the wave functions, 
implies a strong modification on the contribution of the large dipole sizes, as observed in Fig. \ref{fig1}. It is important
to emphasize that this contribution dominates the cross section, i.e. if we disregard the saturation effects, the diffractive cross section is dominated by soft physics.

\begin{figure}
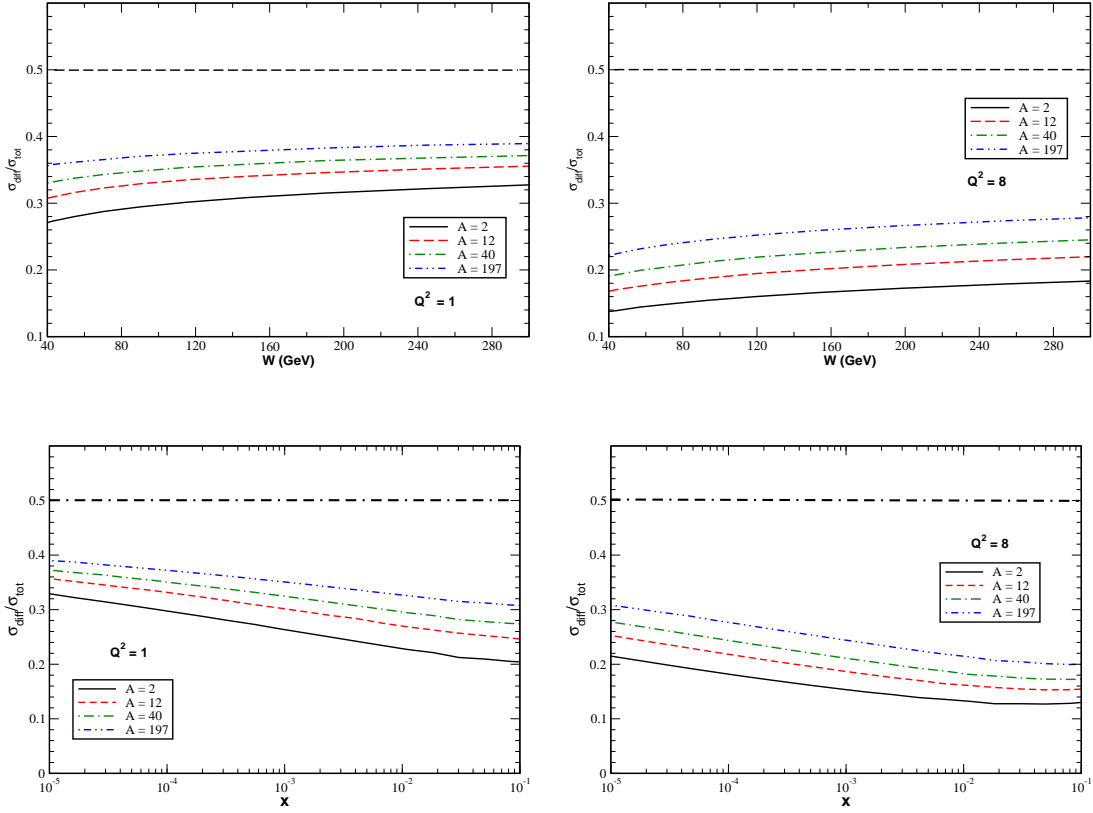

\centerline{
\begin{tabular}{ccc}
{\psfig{figure=fig2a.eps,width=7.cm}} & \,\,\, {\psfig{figure=fig2b.eps,width=7.cm}} &  \\
&  & \\
{\psfig{figure=fig2c.eps,width=7.cm}} & \,\,\,  {\psfig{figure=fig2d.eps,width=7.cm}} & \\
 &  & \\
\end{tabular}}
\caption{The ratio between the  diffractive and total  cross section as a function of $x$ and $W$ for different values 
of $A$ and $Q^2$. The black disk limit, $\sigma_{diff}$/$\sigma_{tot} = 1/2$,  is also presented. }
\label{fig2}
\end{figure}

We now  present a qualitative analysis of the $A$ and energy dependence of the ratio 
$\sigma_{diff}$/$\sigma_{tot}$ using the IIM model generalized for nuclear targets. Following Ref. \cite{GBW} and 
assuming that $\sigma_{dip}$ in the saturation regime can be approximated by $\sigma_0$, 
 the transverse part of  the inclusive and diffractive cross sections, 
in the kinematical range where $Q^2 > Q_s^2$, can be expressed as
\begin{eqnarray}
\sigma_T \approx \int_0^{4/Q^2} \frac{d\rr^2}{\rr^2} \sigma_0 [\frac{\rr^2 Q_s^2}{4}]^{\gamma_{eff}} + \int_{\frac{4}{Q^2}}^{\frac{4}{Q_s^2}} \frac{d\rr^2}{\rr^2} \left(\frac{1}{Q^2 \rr^2}\right) \sigma_0 [\frac{\rr^2 Q_s^2}{4}]^{\gamma_{eff}} + \int_{4/Q_s^2}^{\infty} \frac{d\rr^2}{\rr^2} \left(\frac{1}{Q^2 \rr^2}\right) \sigma_0 \nonumber \\
\end{eqnarray} 
and 
\begin{eqnarray}
\sigma_T^D \approx \frac{1}{B_A} \left[\int_0^{4/Q^2} \frac{d\rr^2}{\rr^2} \sigma_0^2 [\frac{\rr^2 Q_s^2}{4}]^{2\gamma_{eff}} + \int_{\frac{4}{Q^2}}^{\frac{4}{Q_s^2}} \frac{d\rr^2}{\rr^2} \left(\frac{1}{Q^2 \rr^2}\right) \sigma_0^2 [\frac{\rr^2 Q_s^2}{4}]^{2\gamma_{eff}} + \int_{4/Q_s^2}^{\infty} \frac{d\rr^2}{\rr^2} \left(\frac{1}{Q^2 \rr^2}\right) \sigma_0^2 \right]\,. \nonumber \\
\end{eqnarray} 
In order to obtain an approximated expression for the ratio we will disregard the $\rr$-dependence of the effective 
anomalous dimension, i.e. $\gamma_{eff} = \gamma = \mbox{cte}$. In this case, we obtain $\sigma_{diff}/\sigma_{tot} \approx [\frac{Q_s^2}{Q^2}]^{1-\gamma}$. 
Assuming $\gamma = 0.84$, as in Ref. \cite{IIM}, we predict that the ratio decreases with the photon virtuality and 
presents a weak energy dependence. However, analyzing the $A$-dependence, we expect a growth of approximately 30 $\%$ when we increase $A$ from 2 to 197.
In the kinematical range where $Q^2 < Q_s^2$ the ratio of cross sections presents a similar behavior. The main 
difference is that in the asymptotic regime of very large energies the cross section for diffraction reaches the 
black disk limit of $50 \%$ of the total cross section.    

In Fig.  \ref{fig2} we show the ratio
$\sigma_{diff}$/$\sigma_{tot}$, computed  with  the help of (\ref{sig_gp}) and 
(\ref{sig_difra}),  as a function of $W$ and $x$ for different values of $A$. The black disk limit, $\sigma_{diff}$/$\sigma_{tot} = 1/2$, is also presented in the figure. 
We can see that the ratio  depends  weakly on $W$ and on $x$ but is strongly 
suppressed for increasing $Q^2$. This suggests that in the deep perturbative region, 
diffraction is more suppressed. This same behavior was observed in diffractive $ep$ 
data \cite{e_p_data}.  Moreover, the energy dependence of the ratio is remarkably flat, 
increasing with  $A$, becoming 37 \% (30 \%) larger for gold  
in comparison to proton (deuteron).  This behavior agrees qualitatively with the previous 
calculation of \cite{levlub} and with our previous estimate.  Similar results have been obtained  in the pioneering work of  Ref. \cite{nzz} in a different context. The appearance of a large rapidity gap in 37 \% of all $eA$ scattering events would be a striking confirmation 
of the saturation picture.

\begin{figure}
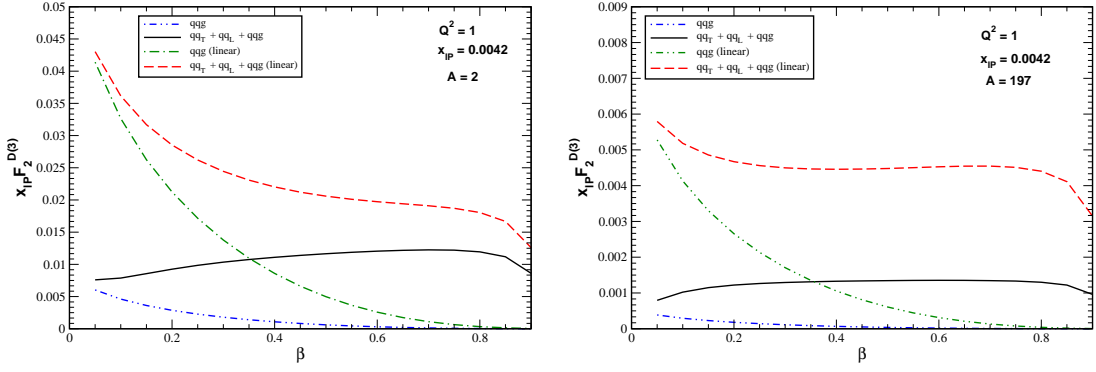

\centerline{
\begin{tabular}{ccc}
{\psfig{figure=fig3a.eps,width=7.cm}} & \,\,\, {\psfig{figure=fig3b.eps,width=7.cm}} &  \\
 &  & \\
\end{tabular}}
\caption{Diffractive structure function $F_2^{D(3)}$ as a function of $\beta$ and distinct nuclei. The $q\bar{q}g$  component of the diffractive structure function is explicitly  presented.}
\label{fig3}
\end{figure}

\begin{figure}
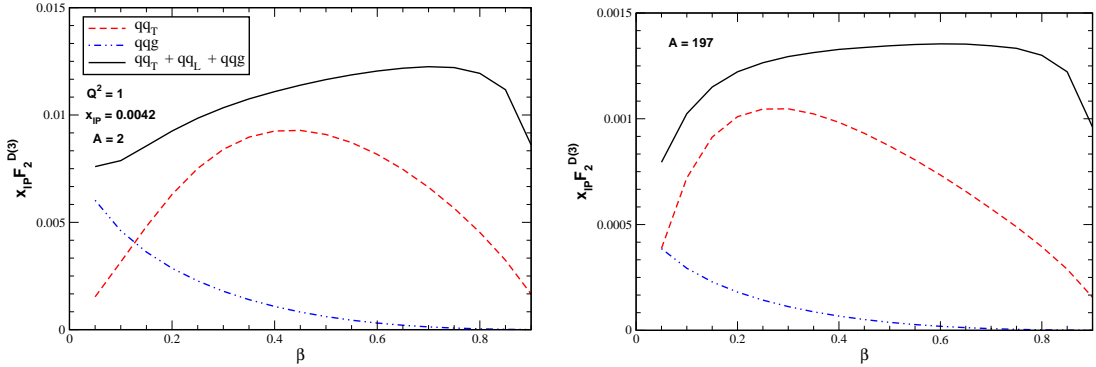

\vspace*{1.5cm}
\centerline{
\begin{tabular}{ccc}
{\psfig{figure=fig4a.eps,width=7.cm}} & \,\,\, {\psfig{figure=fig4b.eps,width=7.cm}} &  \\
 &  & \\
\end{tabular}}
\caption{Diffractive structure function $F_2^{D(3)}$ as a function of $\beta$ and distinct nuclei. The transverse and $q\bar{q}g$  components of the diffractive structure function are explicitly  presented.}
\label{fig4}
\end{figure}

In Fig. \ref{fig3}  we show our predictions for  the diffractive 
structure functions $x_{I\!\!P} F_2 ^{D(3)}(x_{I\!\!P}, \beta, Q^2)$ as a function of $\beta$ and different nuclei.
We also present the linear prediction for $x_{I\!\!P} F_2 ^{D(3)}$. It is 
important to emphasize that a linear ansatz for the dipole cross section would not describe the HERA data. However, in order
to estimate the importance of the saturation physics and clarify its contribution at different kinematical ranges, a comparison between these
two predictions is valid. We can see that  the normalization of $x_{I\!\!P} F_2 ^{D(3)}$ is strongly reduced 
increasing the atomic number, what is 
expected from our analysis of the diffractive overlap function. Moreover, although the photon wavefunction 
determines the general structure of the $\beta$-spectrum \cite{wusthoff,GBW}, the  $q\bar{q}g$ component, which 
dominates the region of small $\beta$, has its behavior modified 
by  saturation effects and changes the  behavior of  $x_{I\!\!P} F_2 ^{D(3)}$ in this region. Moreover, the diffractive structure function becomes almost flat at intermediate values of 
$\beta$ and large $A$. 
In Fig. \ref{fig4} we show an amplification of the lower curves in Fig. \ref{fig3} and include also the $q\bar{q}_T$ component. In doing this another interesting feature of diffraction off nuclear targets emerges, namely, the relative reduction of
the $q\bar{q}g$ component with respect to the  $q\bar{q}$ one.

\begin{figure}
\vspace*{1.5cm}
\centerline{
{\psfig{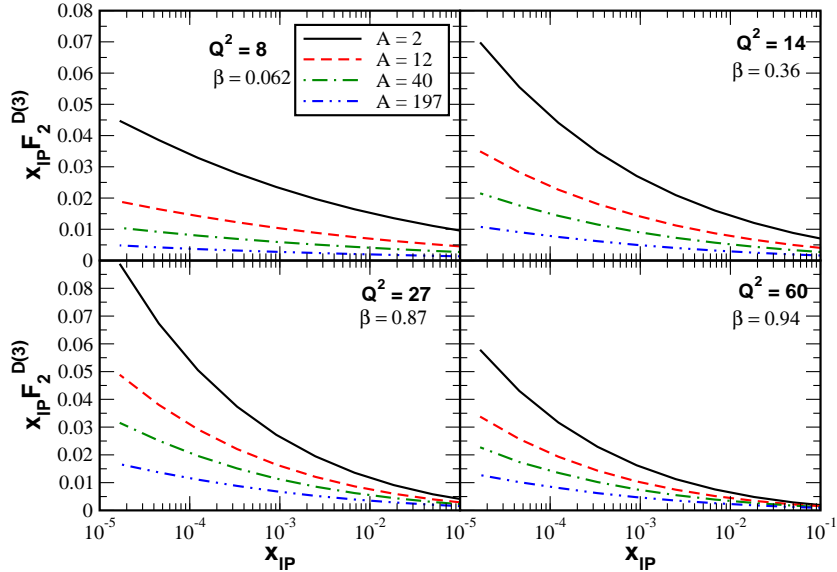}}}
\caption{Predictions for the diffractive structure functions $x_{I\!\!P} F_2 ^{D(3)}(x_{I\!\!P}, \beta, Q^2)$ plotted as a function of $x_{I\!\!P}$ for different values of $\beta$, $Q^2$ and $A$.}
\label{fig5}
\end{figure}

In Fig. \ref{fig5} we show our predictions for  
$x_{I\!\!P} F_2 ^{D(3)}(x_{I\!\!P}, \beta, Q^2)$ as a function of $x_{\pom}$ and different values of $\beta$, $Q^2$ 
and $A$. Our choice for the combination of values of $\beta$ and $Q^2$ was motivated by the HERA results 
\cite{e_p_data}. The $x_{\pom}$ dependence comes from the dipole cross section, which in our case is given by the 
IIM model generalized to nuclear targets.
We find that $x_{I\!\!P} F_2 ^{D(3)}$ increases at small values of $x_{\pom}$. However, as the saturation scale grows with $A$, the $x_{\pom}$ becomes weaker  when we increase the atomic number.

%

\begin{figure}
\centerline{
{\psfig{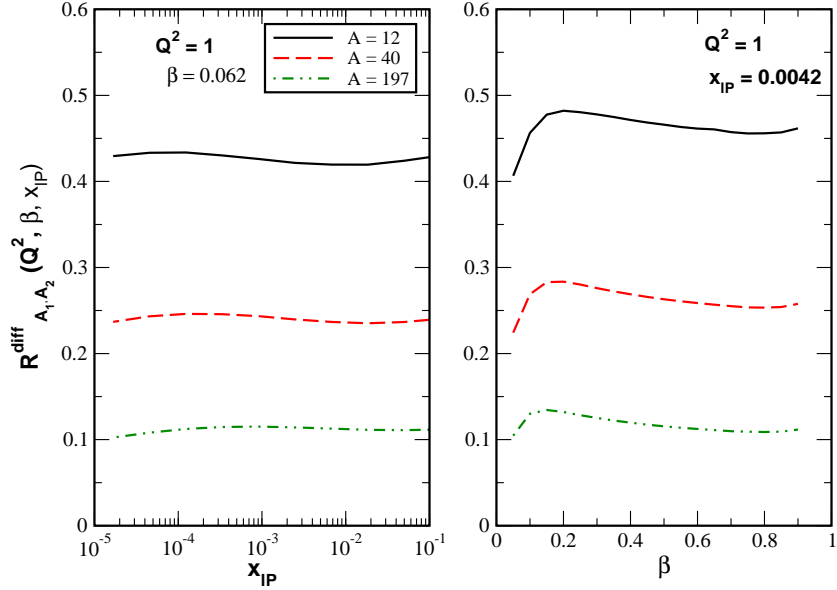}}}
\caption{The ratio $R^{diff}_{A1,A2}(Q^2, \beta, x_{I\!\!P}) = \frac{F_{2,A1}^{D(3)}(Q^2,\beta, x_{I\!\!P})}{F_{2,A2} ^{D(3)}(Q^2, \beta, x_{I\!\!P})}$ as a function of $x_{\pom}$ and $\beta$. Comparison between the predictions for the ratio at different values of $A_1$.}
\label{fig6}
\end{figure}

\begin{figure}
\vspace*{1.5cm}
\centerline{
{\psfig{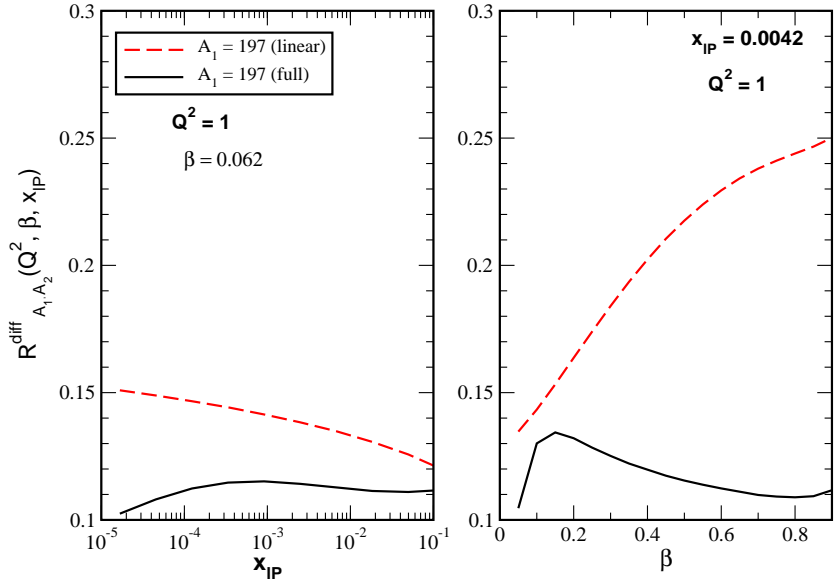}}}
\caption{The ratio $R^{diff}_{A1,A2}(Q^2, \beta, x_{I\!\!P}) = \frac{F_{2,A1} ^{D(3)}(Q^2,\beta, x_{I\!\!P})}{F_{2,A2} ^{D(3)}(Q^2,\beta, x_{I\!\!P})}$ as a function of $x_{\pom}$ and $\beta$. Comparison between the saturation and linear predictions.}
\label{fig7}
\end{figure}

In Fig. \ref{fig6} we show our predictions for the ratio 
$$
R^{diff}_{A1,A2}(\beta, Q^2, x_{I\!\!P}) = \frac{F_{2,A1} ^{D(3)}(\beta, Q^2, x_{I\!\!P})}
{F_{2,A2} ^{D(3)}(\beta, Q^2, x_{I\!\!P})}
$$ 
as a function of $\beta$ and $x_{I\!\!P}$. In our calculation we assume that $A_2 = 2$. Our analysis is motivated by 
Refs. \cite{Arneodoetal, raju_eRHIC}. In these papers it was suggested that  the nuclear dependence of this ratio
can help us to establish  the universality of the 
Pomeron structure.  In particular, in Ref. \cite{Arneodoetal}, the authors have pointed out that if $R^{diff}_{A1,A2}(\beta, Q^2, x_{I\!\!P})=1$ one can conclude that the structure of the Pomeron is universal  and the Pomeron flux is $A$ independent. On the other hand, if  $R^{diff}_{A1,A2}(\beta, Q^2, x_{I\!\!P})= f (A_1,A_2)$, the structure is universal but the flux is $A$ dependent. From our previous analysis we can anticipate 
that in the dipole picture, assuming the presence of saturation effects,  this ratio will be $A$ dependent. Moreover, its behavior  will be determined by the saturation scale.
 In    Fig. \ref{fig6} we observe a strong decrease of $R^{diff}_{A1,A2}$ as a function of $A$. 
At the same time this ratio is remarkably flat at all values of $A$, $x_{\pom}$ and $\beta \ge 0.2$. However, at small $\beta$, it presents a steeper dependence directly associated with the  nuclear dependence of the $q\overline{q}g$ component of the diffractive structure function (See Fig. \ref{fig4}). 
In order to estimate how much of the flat behavior is due to saturation we calculate $R^{diff}_{A1,A2}$  
again using only the linear part of the dipole cross section, as discussed above, for the heaviest target ($A = 197$), 
for which the saturation effects are expected to be dominant and show our results in Fig. \ref{fig7}.
As it can be seen, saturation is largely responsible for the weak dependence of $R^{diff}_{A1,A2}$ on $x_{\pom}$ and $\beta$.  
 In Ref. \cite{Arneodoetal} the possibility that the $A$ dependence of $R^{diff}_{A1,A2}$ can be described by the ratio between the inclusive nuclear structure functions was suggested  for the case of an $A$ dependent Pomeron structure and  $A$ independent  flux. 
 We have checked this conjecture using the results from Ref. \cite{eA} and have found that it fails, the inclusive ratio
 being larger than the diffractive one.

\section{Summary}
\label{conc}

Diffractive physics in nuclear DIS experiments could be studied at the electron-ion collider 
eRHIC. Hence it is interesting to extend the current $ep$ predictions  to  the corresponding   energy and targets which will be  available in this collider.
In this work we address nuclear diffractive DIS and compute observable quantities 
like $\sigma_{diff}$/$\sigma_{tot}$ and  $F_{2}^{D(3)}$ in the dipole picture. In particular,  we have 
investigated the potential of $eA$ collisions as a tool for revealing the details 
of the saturation regime.  Since $\sigma_{diff}$ is proportional to $\sigma_{dip}^2$, diffractive 
processes are expected to be particularly sensitive to saturation effects. Moreover,
due to the highly non-trivial $A$ dependence of  $\sigma_{dip}$,  diffraction off nuclear 
targets is even more sensitive to non-linear effects. Using  well established 
definitions of $\sigma_{diff}$ and $F_2 ^{D(3)}$  and a recent and successful 
parametrization of $\sigma_{dip}$,  we have studied observables which may serve as 
signatures of the Color Glass Condensate. Without adjusting any parameter, we have found that the ratio   
$\sigma_{diff}$/$\sigma_{tot}$ is a very flat function of the  center-of-mass energy $W$, in good agreement with existing HERA data. Extending the
calculation to nuclear targets, we have shown that this ratio  remains flat and increases
with the atomic number. At larger nuclei we predict that approximately 37 $\%$ of the events observed at 
eRHIC should 
be diffractive. Moreover, we have analyzed the behavior of the diffractive structure function $F_2 ^{D(3)}$ and 
found out that in certain regions of the $\beta$ - $x_{I\!\!P}$ 
space,  the diffractive structure function $F_2 ^{D(3)}$ decreases up to an order of magnitude  
when going from the lightest to heaviest targets.  Finally, we have found that, for 
nuclear targets, the contribution of the  $q\overline{q} g$ Fock state becomes less 
important.

Considering the results obtained in this paper and those presented in Ref. \cite{eA}, we can  conclude that $eA$ collisions are very promising for the experimental 
confirmation of the  non-linear effects of QCD.

\begin{acknowledgments}
VPG thanks Magno Machado for informative and helpful  discussions.   This work was  partially 
financed by the Brazilian funding
agencies CNPq, FAPESP and FAPERGS.
\end{acknowledgments}



\end{document}